\documentclass[a4paper]{article}
\usepackage{multirow}
\usepackage{ISCSLP2024}
\usepackage{hyperref}

\title{CUSIDE-T: Chunking, Simulating Future and Decoding for Transducer based Streaming ASR}

\name{
Wenbo Zhao$^{1}$, Ziwei Li$^2$, Chuan Yu$^1$, Zhijian Ou$^{2}$
\thanks{
This work is partly supported by Guangxi Science and Technology Project (GUIKEAD23026054).
}
}
\address{
  $^1$China Unicom (Guangdong) Industrial Internet Co., Ltd., China\\
  $^2$Speech Processing and Machine Intelligence (SPMI) Lab, Tsinghua University, China 
}
\email{
zhaowb19@chinaunicom.cn, ozj@tsinghua.edu.cn
}

\begin{document}

\maketitle
\begin{abstract}
Streaming automatic speech recognition (ASR) is very important for many real-world ASR applications. However, a notable challenge for streaming ASR systems lies in balancing operational performance against latency constraint.
Recently, a method of chunking, simulating future context and decoding, called CUSIDE, has been proposed for connectionist temporal classification (CTC) based streaming ASR, which obtains a good balance between reduced latency and high recognition accuracy.
In this paper, we present CUSIDE-T, which successfully adapts the CUSIDE method over the recurrent neural network transducer (RNN-T) ASR architecture, instead of being based on the CTC architecture.
We also incorporate language model rescoring in CUSIDE-T to further enhance accuracy, while only bringing a small additional latency.
Extensive experiments are conducted over the AISHELL-1, WenetSpeech and SpeechIO datasets, comparing CUSIDE-T and U2++ (both based on RNN-T).
U2++ is an existing counterpart of chunk based streaming ASR method.
It is shown that CUSIDE-T achieves superior accuracy performance for streaming ASR, with equal settings of latency.
\end{abstract}

\section{Introduction}
% Recently, automatic speech recognition (ASR) system can be divided into two main categories, namely traditional DNN-HMM hybrid systems \cite{chen2016training} and end-to-end (E2E) systems. The hybrid system was abandoned due to its complex structure and low efficiency. 
Streaming ASR (also known as online ASR) is very important for many real-world applications, whose goal is to emit recognition results as quickly and accurately as possible on the fly when the user is speaking.
Recently, streaming methods have been mainly studied for end-to-end ASR.
% In contrast, E2E systems not only extremely simplified training pipeline but also show superior performance in the standard of word error rate (WER). 
There exist three end-to-end (E2E) approaches: connectionist temporal classification (CTC) \cite{graves2006connectionist}, recurrent neural network transducer (RNN-T) \cite{graves2012sequence}, and attention-based encoder-decoder (AED) \cite{chorowski2014end,chan2015listen,chorowski2015attention}. 
While it is common for all these approaches to define a differentiable sequence-level loss of mapping the acoustic sequence to the label sequence, 
the three approaches do have different properties, especially for streaming ASR.
% They have advantages and limitations in recognition accuracy and speed, and many efforts have been paid for identifying the optimal model by comparison of these models \cite{prabhavalkar2017comparison}.

The CTC approach \cite{graves2006connectionist} features an encoder-only simple architecture, which can be trained efficiently and can be easily applied for chunk-based streaming ASR. But it makes the conditional independence assumption between states, which hurts its modeling ability.
% However, the lack of language modeling in this architecture usually leads to a sub-optimal recognition accuracy; 
Both the RNN-T and the AED approaches overcome the conditional independence assumption.
For RNN-T \cite{graves2012sequence}, streaming ASR can be realized based on modification on encoders such as by chunking or using causal neural network based encoders.
For AED \cite{chorowski2014end,chan2015listen,chorowski2015attention}, more efforts are needed to convert full sequence soft attention into local attention in decoder \cite{chiu2017monotonic}, in addition to modifications on encoders.

% the third approach, based on neural transducer loss, namely Recurrent Neural Network Transducer (RNN-T), integrates language models in the E2E model and fits well with the streaming requirement, therefore it has been widely adopted.

% In these end-to-end (E2E) systems, the acoustic model often adopts Transformer \cite{miao2020transformer} and Conformer \cite{gulati2020conformer} architectures. However, the full sequence attention mechanism involves sequence-level fully connected computation,and not suitable for streaming ASR. For streaming ASR, chunk-based self-attention networks, which use chunk-level input to control the latency. 

Chunk-based methods have been widely used for modifying the encoders for streaming ASR for all the three E2E approaches.
In chunk-based methods, a certain number of history and future frames are usually spliced into each chunk, which benefits the acoustic encoder and improves ASR performance. While the use of history frames is straightforward, we need to wait for the real future frames, which brings non-trivial latency. The multi-mode Transformer Transducer \cite{kim2021multi} proposes to consider stochastic future contexts during training, so that the trained model can adapt to varying latency constraints in different scenarios. 
% However, which does not resolve the latency issue stemming from future information, and may lead to significant accuracy deterioration  when aiming for low latency settings. 
% A recently proposed new framework CUSIDE \cite{an2022cuside} by a new simulation module is introduced to recursively simulate the future contextual frames, without waiting for future context.
Recently, a method of \underline{{C}}h\underline{{u}}nking, \underline{{Si}}mulating Future Context and \underline{{De}}coding, called CUSIDE, is proposed \cite{an2022cuside}, which introduces a new simulation module to recursively simulate the future contextual frames, without waiting for future context, and obtains state-of-the-art streaming ASR results.

The original CUSIDE method is built upon the CTC \cite{graves2006connectionist} and CTC-CRF \cite{CTCCRF_IC19} models.
In this work, we propose CUSIDE-T which incorporates the CUSIDE methodology into the RNN-T based model for streaming ASR. 
An important feature is that CUSIDE-T introduces a simulation module to simulate future context. The simulation module is lightweight and fast, and can be jointly optimized with RNN-T from scratch, without much additional computation overhead. 
Inheriting from CUSIDE, other features of CUSIDE-T include joint training of streaming and non-streaming unified models, chunk jittering and right context randomization in training.
Moreover, we use a language
model (LM) rescoring decoding strategy with CUSIDE-T.

An existing counterpart of chunk based streaming ASR method is U2++ \cite{wu2021u2++}. U2 is a unified streaming and non-streaming two-pass model for speech recognition and adopts the hybrid CTC/attention approach \cite{zhang2020unified}.
In the two pass decoding, a CTC decoder outputs first pass hypotheses in a streaming way. At the end of the input, an left-to-right attention decoder uses full context attention to get better results. In U2++, another right-to-left attention decoder is further added. U2 uses dynamic chunk training, similar to CUSIDE and also the multi-mode Transformer Transducer \cite{kim2021multi}, which use stochastic future contexts during training.
% The dynamic chunk size follows a uniform distribution from 1 to the utterance length.

CUSIDE-T and U2++ have some common features such as joint training of streaming and non-streaming unified models, and chunk size randomization. An important difference between CUSIDE-T and U2++ is that CUSIDE-T uses future context simulation. In this paper, we compare CUSIDE-T and U2++, both within the RNN-T approach, i.e., the U2++ model and the CUSIDE-T model use the same RNN-T architecture.
Experimental results on the AISHELL-1 \cite{bu2017aishell} dataset, the WenetSpeech \cite{zhang2022wenetspeech} dataset, and the SpeechIO benchmarking demonstrate that CUSIDE-T achieves superior accuracy performance for streaming ASR, with equal settings of latency, revealing the advantage of future context simulation.
We further find that U2++ with LM rescoring is competitive to U2++ with attention rescoring, and CUSIDE-T with LM rescoring produces the best results.

% All comparisons are made in the standard RNN-T architecture, thus the works like \cite{variani2020hybrid} that modify the architecture or training objective of RNN-T are not included in this work. 

\section{Background and Related Work}

\subsection{RNN-T model}

The RNN-T model is a popular choice for streaming ASR applications with the state-of-the-art recognition performance. Given a sequence of acoustic feature $\mathbf{x}=(x_{1},x_{2},...,x_{T})$ and the corresponding label sequence $\mathbf{y}=(y_{1},y_{2},...,y_{U})$, the model estimates a conditional probability distribution $P(\mathbf{y}|\mathbf{x})$. 
A typical RNN-T model comprises three components: an encoder, a predictor, and a joint network. The encoder network converts $\mathbf{x}$ into a sequence of high-level representations, called hidden states, $\mathbf{h} = (h_{1},h_{2},...,h_{T})$, denoted by $\mathbf{h}=\text{Encoder}(\mathbf{x})$. The prediction network encodes historical sequence $(y_{1},...,y_{u-1})$ into 
 ${g}_{u}=\text{Predictor}(y_{1},...,y_{u-1})$
. The joint network is a feed-forward network, $\text{Joiner}({h}_{t},{g}_{u})$, taking the outputs from both encoder and prediction network and calculating the probability distribution over output labels, $P(\hat{y}_i|t,u)$, by a softmax function.
The loss function of an RNN-T model is defined as the negative log-likelihood:
\begin{displaymath}
\mathcal{L}=-\log P(\mathbf{y}|\mathbf{x})
\end{displaymath}
where 
\begin{displaymath}
P(\mathbf{y}|\mathbf{x}) = \sum_{\mathbf{\hat{y}} \in \mathcal{A}(\mathbf{x},\mathbf{y}) } \prod_{i=1}^{T+U} P(\hat{y}_i|t_i,u_i),
\end{displaymath}
and $\mathbf{\hat{y}}=(\hat{y}_{1},...,\hat{y}_{T+U}) \in \mathcal{A}(\mathbf{x},\mathbf{y}) $ are alignment sequences with $T$ blanks and $U$ labels such that removing the blanks in $\mathbf{\hat{y}}$ yields $\mathbf{y}$.
% $\mathcal{B}$ is all the possible alignments including blank labels between features $\mathbf{h}$ and the target label sequence $\mathbf{y}$.

\subsection{Chunk-based training}
\label{sec:related-work:chunking-based training}

With the streaming-friendly nature of RNN-T, streaming and non-streaming modes only differ in the acoustic encoder networks. Conventional uni-directional recurrent neural networks like LSTM \cite{hochreiter1997long} are able to perform forward pass in a strictly streaming mode. However, they are not widely used in recent end-to-end ASR systems due to their higher error rates \cite{narayanan2021cascaded, gulati2020conformer}, compared to attention-based encoders like Transformer \cite{vaswani2017attention} and Conformer \cite{gulati2020conformer}. It is challenging to support streaming in attention-based models, which benefit from the full-context modeling. To make a trade-off between recognition accuracy and streaming latency, many previous works use chunk-based training instead \cite{an2022cuside, zhang2020unified, wu2021u2++, dong2019self, miao2020transformer}, i.e., restricting the attention layers to capture a limited range of context (the chunk). In implementation, chunking-based training can be simply done via masking the attention matrix or splicing input utterances to chunking clips.

\subsection{Unified streaming and non-streaming recognition}
\label{sec:related-work:unified-training}

Combining the non-streaming model with full-context in streaming model training has been found to be beneficial for streaming ASR in recent works \cite{zhang2020unified, yu2021dual, narayanan2021cascaded}. The dual-mode ASR \cite{yu2021dual} and U2 \cite{zhang2020unified} propose to share parameters across streaming and non-streaming model, which significantly reduces the parameters and simplifies the workflow of training streaming model. Specifically, the dual-mode ASR conducts a multi-task training manner, where there are a streaming loss and a full-context non-streaming one. U2 unifies the two modes through a dynamic chunking strategy, that is, uniformly drawing a chunk size from a minimum value to full length of the utterance during training. With the parameter sharing strategy, the dual-mode ASR and U2 have shown to perform well on both streaming and non-streaming inference scenarios.
% Interestingly, the dual-mode ASR could obtain slightly better WER in non-streaming test compared to single-mode non-streaming model

\subsection{Decoding in streaming ASR}

For naturally streaming models like CTC and RNN-T, the frame-synchronous decoding is fully streaming as long as the encoders are causal or chunking-based. In non-streaming inference, it is useful to combine contextual information into the decoding via language model (LM) integration \cite{hannun2014deep, mikolov2010recurrent} or attention rescoring \cite{watanabe2017hybrid, zhang2020unified}. 
These two techniques generally further improves the accuracy significantly.
It is feasible to incorporate the contextual information in a streaming system in one-pass decoding; however, the real time factor (RTF) might deteriorate due to the redundant computational overhead introduced in beam search \cite{zhang2020unified}. A compromise is to switch to a two-pass decoding manner, i.e., obtaining several hypotheses (a.k.a., the n-best list) from the first-pass decoding and rescoring the hypotheses with the contextual models (LM rescoring or attention rescoring). The computational latency introduced by the second-pass is negligible compared to the auto-regressive beam search process. In rescoring, a linear interpolation between the scores obtained from the first and the second passes is often conducted, for example, in LM rescoring:

\begin{displaymath}
    \mathbf{y}^* = \mathop{\arg\max}_{\mathbf{y}} \left( \lambda_1 \text{Score}_{1}(\mathbf{y}|\mathbf{x}) + \lambda_2 \text{Score}_{2}(\mathbf{y}) \right)
\end{displaymath}

where $\lambda_1$ and $\lambda_2$ are constants tuned on a held-out development set.
Notably, the score in attention rescoring depends on the input $\mathbf{x}$ and should be denoted by $\text{Score}_{2}(\mathbf{y}|\mathbf{x})$.
% However, systematically, the second-pass has to wait for the whole input utterances to get the n-best list, the decoding is therefore not streaming in the strict sense.

%\section{CUSIDE-T}

\begin{figure}[t]
  \centering
  \includegraphics[width=\linewidth]{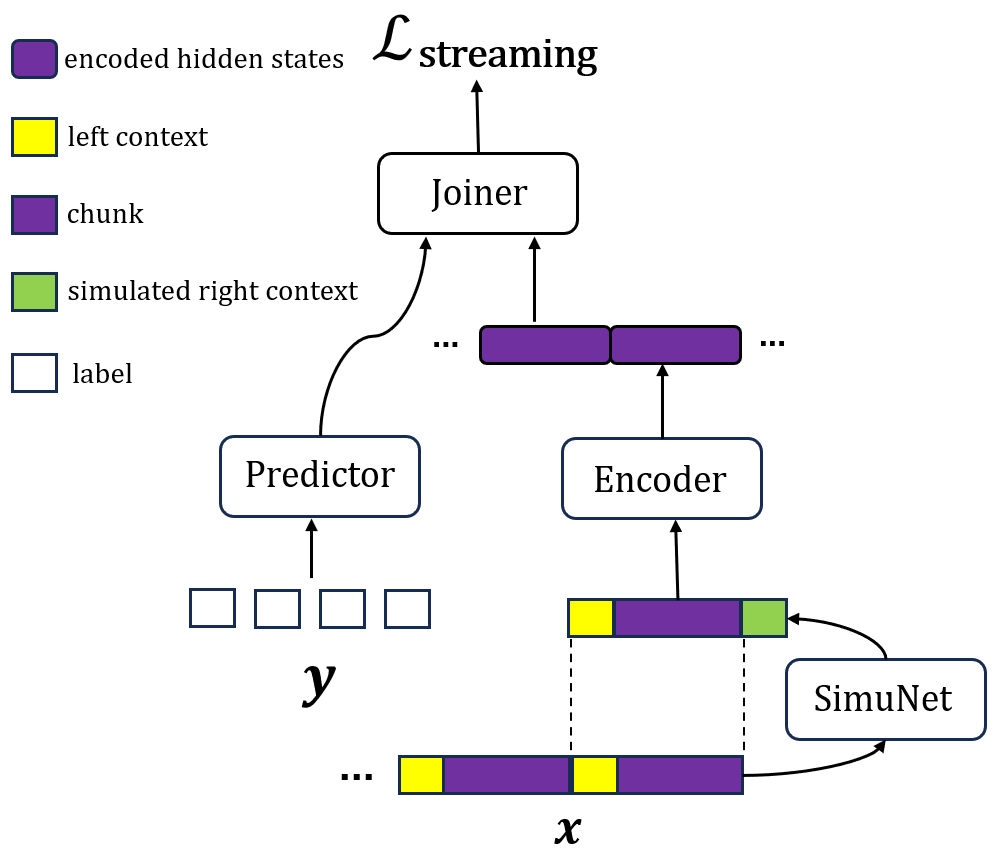}
  \caption{Overview of CUSIDE-T. The encoder, predictor and joiner are the network components of RNN-T. Input spectrum features are first split into overlapped chunks, then concatenated with a simulated right context predicted by the simulate network (SimuNet). The target label sequence $\mathbf{y}$ is fed into the predictor. Hidden encoded states, where those from the left and right contexts are eliminated, are spliced with the output of the predictor to fed to the joiner to calculate the conditional probability of labels.}
  \label{fig:cuside-t}
\vspace{-4mm}
\end{figure}

\section{CUSIDE-T}
In this section, we give details of our design of CUSIDE-T, which incorporates CUSIDE within RNN-T for streaming ASR and uses external language model fusion. Overview of CUSIDE-T is shown in  Figure \ref{fig:cuside-t}.

\subsection{Chunking and context simulation}
\label{sec:cuside-t:simu}

As discussed in Section \ref{sec:related-work:chunking-based training}, chunking-based training is common to utilize non-causal networks in streaming ASR. In our experiments, the chunk size is drawn from a uniform distribution $\text{Unif}(C-A, C+A)$, where $C$ is the default size and $A$ is the dynamic range. To further improve the recognition accuracy, some extra frames are concatenated with current chunk, which are denoted as ``left context" and ``right context" respectively, which is called context sensitive chunking (CSC) \cite{an2022cuside}.
% Besides the extra computation with the overlapped frames, 
Note that when we include any right context frames in CSC, the inference process has to wait for these future frames and will suffer higher latency. Instead, CUSIDE proposes to simulate the right context through an extra simulated network (SimuNet) based on existing historical frames. Specifically, in our implementation, a SimuNet consists of a GRU and a feed-forward layer. Historical frames are fed into SimuNet recursively, and the simulated right contextual frames for the current chunk is parallelly obtained via linear transformations. In training of SimuNet, a L1 loss between the simulated frames and real frames is used during training. It should be noted that, though SimuNet brings some extra parameters and computation, they are negligible compare to those of the encoder. We will discuss this later in Section \ref{sec:exp}.

\subsection{Multi-objective training}

Previous works have shown that the streaming model can benefit from joint non-streaming training, as discussed in Section \ref{sec:related-work:unified-training}. Here we use a similar multi-objective training (MoT) strategy as in CUSIDE \cite{an2022cuside}, i.e., training the model with a streaming RNN-T loss $\mathcal{L}_\text{streaming}$, a non-streaming RNN-T loss $\mathcal{L}_\text{non-streaming}$, and a simulation loss. The final training loss in CUSIDE-T is as follows, which involves two passes of forward calculations, one for $\mathcal{L}_\text{streaming}$, and the other for $\mathcal{L}_\text{non-streaming}$:
\begin{displaymath}
    \mathcal{L}_\text{MoT} = \mathcal{L}_\text{streaming} + \mathcal{L}_\text{non-streaming} + \alpha \mathcal{L}_\text{simu}
\end{displaymath}
where $\mathcal{L}_\text{simu}$ is the L1 loss of the SimuNet mentioned in Section \ref{sec:cuside-t:simu} and $\alpha$ is a weight factor. 
% It is known that RNN-T training is somewhat memory-demanding, and futher considering the extra computation brought by MoT training and overlapped chunk, it is expected that CUSIDE-T training would be slower than a standard RNN-T.

\subsection{External language model fusion}

Language model integration \cite{zheng2022empirical} adopts an external language model (ELM) that can be trained using additional text data to improve the language modeling ability of an ASR model. 
A number of fusion approaches have been proposed to address the problem of LM integration for E2E models, among which the shallow fusion is the most popular one. 
Shallow fusion conducts log-linear interpolation between the scores of the E2E model and the ELM as follows:
% It adds up the posterior probability from RNN-T and the language model probability in the logarithmic domain and introduces an extra parameter $\beta$ for the length penalty:
\begin{displaymath}
    \log P_{\text{RNN-T}}(\mathbf{y}|\mathbf{x})+\lambda\log P_{\mathrm{ELM}}(\mathbf{y})+\beta|\mathbf{y}|
\end{displaymath}
where $P_{\text{RNN-T}}(\mathbf{y}|\mathbf{y})$ is the posterior probability from RNN-T, and $\beta$ represents weighting reward for the output sequence length.

\section{Experiments}

\subsection{Datasets}

In our experiments, the CUSIDE-T models are trained on two Chinese Mandarin labeled speech datasets respectively: the AISHELL-1 dataset \cite{bu2017aishell} and the WenetSpeech dataset \cite{zhang2022wenetspeech}.
The AISHELL-1 contains a 150-hour training set, a 20 hour development set, and a 10-hour test set. The test set contains 7176 utterances in total. WenetSpeech is a large dataset of industrial scale in fields such as entertainment, journalism, literature, technology and free conversation and so on. It is collected from online websites YouTube and Podcast and transcribed through optical character recognition (OCR) and ASR and contains a 10,000-hour training set, and we use $dev$ as development set and $test\_net$ , $test\_meeting$ that contains about 33,100 utterances in total as the test set for evaluation. Further, a benchmarking dataset SpeechIO\footnote{https://github.com/SpeechColab/Leaderboard} is used for additional evaluation. The SpeechIO testsets consist of 18 datasets SpeechIO-\{01-18\} collected from various scenarios.

\subsection{Experimental Setup}
\label{sec:exp}

We use CAT \cite{CTCCRF_IC19,CAT}, an end-to-end open-source speech recognition toolkit, to conduct our experiments.
For all experiments, the speech features are 80-dimensional
log Mel-filterbank (FBank) computed by Kaldi\footnote{https://github.com/kaldi-asr/kaldi} with a 25ms window and a 10ms shift. Mean and variance normalization are applied to the features. We also applied SpecAugment \cite{park2019specaugment} for acoustic data augmentation for all experiments. The 3-fold speed perturbation is used on AISHELL-1 experiments for data augmentation, but not on WenetSpeech.
The modeling units are Chinese characters of around 4k size in AISHELL-1 and around 5.5K in WenetSpeech experiments.
% and all experiments are conducted using PyTorch. 

Adam optimizer with Transformer learning rate scheduler \cite{vaswani2017attention} is used with learning rate at 0.0003 and gradient clipping at 5.0 for all models, and the learning rate is warmed up with 25,000 steps.
The total training iterations is 95,000 for AISHELL-1 and 150,000 for WenetSpeech.
% We stop training when the iteration steps are less than a given threshold. 
In WenetSpeech experiments, a 2-fold gradient accumulation is used. At the end of training, 10 models with the best losses on the dev set are averaged to produce the final checkpoint for evaluation. The beam size of the beam search is 16.

The RNN-T architectures used in our experiments for both CUSIDE-T and U2++ are identical, i.e., Conformer-based Transducer of around 90M parameters, where the encoder network consists of a 2-layer VGG for 1/4 subsampling and a 12-layer Conformer with 4 attention heads, 512 attention dimensions and 2048 feed-forward units. The prediction network is a single layer LSTM and the joint network is a fully-connected layer with tanh activation and 512 hidden units. The SimuNet, described in Section \ref{sec:cuside-t:simu}, is a 3-layer GRU with 256 hidden units and followed by a fully-connected layer, being about 5$\%$ of the total model size.

For inference, %the standard RNN-T decoding algorithm in \cite{graves2012sequence} is not used, instead, 
we adopt the monotonic RNN-T (also known as recurrent neural aligner, RNA) decoding algorithm in \cite{tripathi2019monotonic} for faster beam search. The decoding process runs in two-pass manner. At the first pass, the standalone RNN-T model using beam search of width 16 obtains hypotheses of n-best. At the second pass, we apply attention rescoring or LM rescoring for U2++ and LM rescoring for CUSIDE-T. N-gram LMs in experiments are trained using KenLM toolkit \cite{heafield2011kenlm}.

\subsection{AISHELL-1 Task}

To compare CUSIDE-T and U2++, we first evaluate on the AISHELL-1 dataset.
The released code from U2++ is used, where dynamic chunk-based attention strategy is used in training. 
For decoding with U2++, we test with two settings for the chunk size, 400ms and 640ms, respectively.
For CUSIDE-T, left context size is 800ms for history information, and right context size is 400ms for future information. During training, the chunk size is randomly sampled from $\text{Unif}(C-A, C+A)$. 
We experiment with two settings for $C$, 400ms and 640ms, respectively, both with $A$=200ms.
Speed perturbation with 0.9, 1.0, 1.1 is applied on the whole data.

\begin{table}[t]
\setlength{\arrayrulewidth}{0.5mm}
\setlength{\tabcolsep}{5pt}
\renewcommand{\arraystretch}{1.0}
    \caption{Test set CER ($\%$) of CUSIDE-T and U2++ trained on AISHELL-1 Mandarin speech data. The language model is a word-level 3-gram from training transcript. 
    Following \cite{an2022cuside}, the latency is defined as the chunk size plus additional latency (if any). $\Delta$ denotes the additional latency introduced by rescoring the first-pass hypotheses, typically 50ms $\sim$ 100ms for an utterance. The future context simulation cost takes around 2ms.}
    \label{tab:aishell-1}
    \centering
    \begin{tabular}{c c c c}
    \toprule
    \textbf{Exp} & \textbf{Method} & \textbf{Latency (ms)} & \textbf{CER} \\ 
    \midrule
    \multirow{3}{*}{U2++} & Beam Search & 400 & 6.11 \\
                          & Attention rescoring & 400 + $\Delta$ & 5.98 \\
                          & LM rescoring & 400 + $\Delta$ & 5.74 \\
    \midrule
    \multirow{3}{*}{U2++} & Beam Search & 640 & 5.90 \\
                          & Attention rescoring & 640 + $\Delta$ & 5.61 \\
                          & LM rescoring & 640 + $\Delta$ & 5.47 \\
    \midrule
    \multirow{2}{*}{CUSIDE-T} & Beam Search & 400 + 2 & 6.02 \\
                               & LM rescoring & 400 + 2 + $\Delta$ & \textbf{5.51} \\
    \midrule
    \multirow{2}{*}{CUSIDE-T} & Beam Search & 640 + 2 & 5.85 \\
                               & LM rescoring & 640 + 2 + $\Delta$ & \textbf{5.38} \\
    \bottomrule
    \end{tabular}
\vspace{-4mm}
\end{table}

The performance comparison of U2++ and CUSIDE-T is shown in Table \ref{tab:aishell-1}. In terms of the basic results, CUSIDE-T outperforms U2++ consistently across different latency settings, which reveals the advantage of future context simulation in CUSIDE-T. The superority of CUSIDE-T remains, when both augmented with rescoring.
At the setting of 400ms latency, CUSIDE-T with LM rescoring significantly surpasses U2++ with LM rescoring (5.51\% vs 5.74\%). 
At the setting of 640ms latency, CUSIDE-T also outperforms U2++ (5.38\% vs 5.47\%). 
Additionally, the results suggest that for U2++, LM rescoring performs better than attention rescoring.
% while increased latency generally improves CER, the gains are more substantial when combined with advanced rescoring strategies.

\subsection{WenetSpeech Task}

We further evaluate the two methods of CUSIDE-T and U2++, trained on the larger WenetSpeech dataset, using the same configurations used in AISHELL-1 Task.
% CUSIDE-T was trained for around 25 epochs, which took 7.5 days with 10 GeForce GTX 3090 GPUs in our experiments. 
In addition to the two official in-domain test sets from WenetSpeech, we add a more difficult cross-domain benchmarking test, namely with the SpeechIO datasets.

\begin{table}[t]
\setlength{\tabcolsep}{3pt} % Adjusted column separation
\renewcommand{\arraystretch}{1.2} % Adjusted row height
    \caption{Performance comparison between CUSIDE-T and U2++ trained on 10,000 hour of WenetSpeech. A 6-layer Transformer LM trained on training transcripts is used for rescoring. The chunk size is 400ms.}
    \label{tab:wenetspeech}
    \centering
    \begin{tabular}{l l c c c}
    \toprule
    \textbf{Exp} & \textbf{Method} & \multicolumn{3}{c}{\textbf{CER (\%)}} \\ 
    \cmidrule(lr){3-5}
    & & \textbf{net} & \textbf{meeting} & \textbf{speechio} \\
    \midrule
    \multirow{3}{*}{U2++} & Beam Search & 12.54 & 23.79 & 12.57 \\
                          & Attention rescoring  & 11.43 & 22.11 & 11.99 \\
                          & LM rescoring &  11.29 & 20.75 & 11.37 \\
    \midrule
    \multirow{2}{*}{CUSIDE-T} & Beam Search & 12.41 & 23.15 & 12.47\\
                               & LM rescoring & \textbf{11.11} & \textbf{20.01} & \textbf{11.02}\\
    \bottomrule
    \end{tabular}
\vspace{-4mm}
\end{table}

\begin{figure}[t]
  \centering
  \includegraphics[width=\linewidth]{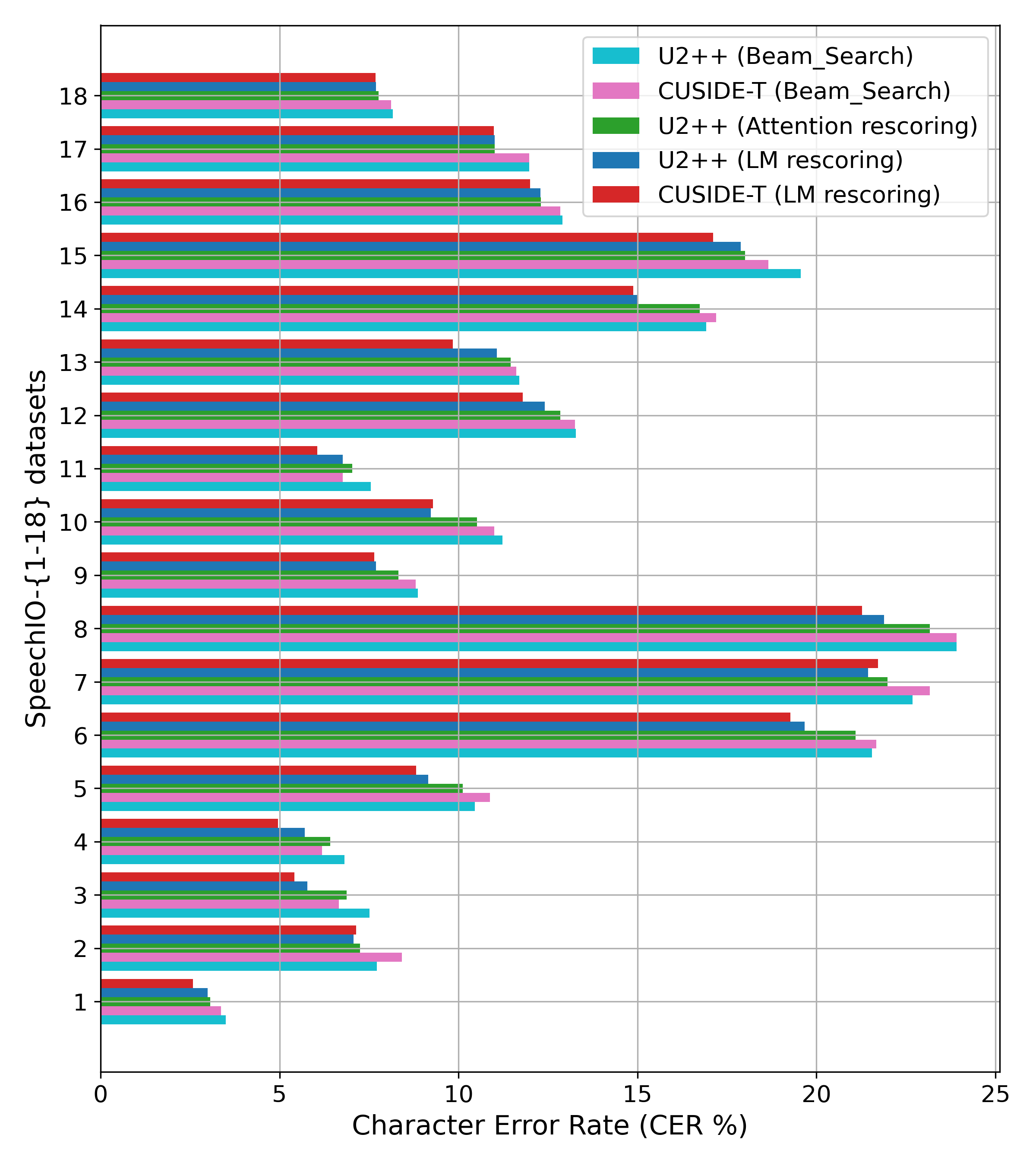}
  \caption{The CER comparison of CUSIDE-T and U2++ on the SpeechIO-\{01-18\} benchmarking data with different decoding manners. The chunk size is 400ms.}
  \label{fig:u2-cuside-t-compared}
\end{figure}

The performance of CUSIDE-T and U2++ is shown in Table \ref{tab:wenetspeech}. 
The chunk size is 400ms in both methods, so with the same latency.
The observation is similar to Table \ref{tab:aishell-1}, i.e., CUSIDE-T performs better than U2++, both in-domain and cross-domain. 
The detailed CERs for the 18 different domains from SpeechIO are shown in Figure \ref{fig:u2-cuside-t-compared}. 
In terms of the basic results, the advantage of CUSIDE-T over U2++ is consistent across different domain, except 4 out of 18 domains. When both augmented with rescoring, CUSIDE-T is boosted more than U2++, surpassing U2++ except 1 out of 18 domains.
% The results show that the CUSIDE-T model performs better on most test sets and performs poorly only on very few datasets, but this defect can be fixed with the help of language models (LM).

\section{Conclusion}

In this paper, we propose CUSIDE-T, which integrates the recent CUSIDE methodology into the RNN-T model for streaming ASR. We present the use of language model rescoring to further improve the performance.
Experiments on both AISHELL-1 and the larger WenetSpeech dataset clearly show that the streaming ASR performance of CUSIDE-T is better than U2++. 
This confirms the advantage of future context simulation in the CUSIDE methodology.
This work also reveals the superiority of LM rescoring over attention rescoring.
% These findings advocate for its adoption in real world applications demanding high precision in speech recognition tasks. This paper is mainly concerned with accuracy. 
% Future works include exploring inference performance optimization of CUSIDE-T.
% The code, scripts and checkpoints used in this paper will be open source for reproduction upon the acceptance of this work.
The code, scripts and checkpoints used in this paper are available at \url{https://github.com/thu-spmi/CAT}.

\bibliographystyle{IEEEtran}

\bibliography{mybib}

% \begin{thebibliography}{9}
% \bibitem[1]{Davis80-COP}
%   S.\ B.\ Davis and P.\ Mermelstein,
%   ``Comparison of parametric representation for monosyllabic word recognition in continuously spoken sentences,''
%   \textit{IEEE Transactions on Acoustics, Speech and Signal Processing}, vol.~28, no.~4, pp.~357--366, 1980.
% \bibitem[2]{Rabiner89-ATO}
%   L.\ R.\ Rabiner,
%   ``A tutorial on hidden Markov models and selected applications in speech recognition,''
%   \textit{Proceedings of the IEEE}, vol.~77, no.~2, pp.~257-286, 1989.
% \bibitem[3]{Hastie09-TEO}
%   T.\ Hastie, R.\ Tibshirani, and J.\ Friedman,
%   \textit{The Elements of Statistical Learning -- Data Mining, Inference, and Prediction}.
%   New York: Springer, 2009.
% \bibitem[4]{YourName17-XXX}
%   F.\ Lastname1, F.\ Lastname2, and F.\ Lastname3,
%   ``Title of your ISCSLP 2024 publication,''
%   in \textit{ISCSLP 2024 -- 23\textsuperscript{rd} Annual Conference of the International Speech Communication Association, September 18-22, Incheon, Korea, Proceedings, Proceedings}, 2024, pp.~100--104.
% \end{thebibliography}

\end{document}